\documentclass{iaus}
\usepackage{graphicx}

\title[W49N OH Masers: anisotropic scattering and Zeeman pairs] 
{OH Maser sources in W49N: probing differential anisotropic scattering with Zeeman pairs}

\author[Avinash A. Deshpande, W. M. Goss \& J. E. Mendoza-Torres]   
{Avinash A. Deshpande$^1$,
W. M. Goss$^2$ \and J. E. Mendoza-Torres$^3$}

\affiliation{$^1$Raman Research Institute, C. V. Raman Avenue, Sadashivanagar, Bangalore, 560080 India\\
email: {\tt desh@rri.res.in} \\[\affilskip]
$^2$National Radio Astronomy Observatory, P.O. Box O, Socorro, NM 87801 USA\\
email: {\tt mgoss@aoc.nrao.edu} \\[\affilskip]
$^3$Instituto Nacional de Astrofisica Optica y Electronica, 72840, Mexico \\
email: {\tt mend@inaoep.mx}}

\pubyear{2012}
\volume{287}  
\pagerange{1--5}
\jname{Cosmic masers - from OH to H0}
\editors{Roy Booth, Liz Humphries \& Wouter Vlemmings, eds.}
\begin{document}

\maketitle

\begin{abstract}
Our analysis of a VLBA 12-hour synthesis observations of the OH masers in W49N has provided
detailed high angular-resolution images of the maser sources, at 1612, 1665 and 1667 MHz.
The images, of several dozens of spots, reveal anisotropic scatter broadening; with typical 
sizes of a few tens of 
milli-arc-seconds and axial ratios between 1.5 to 3. The image position angles oriented perpendicular 
to the galactic plane are interpreted in terms of elongation of electron-density irregularities 
parallel to the galactic plane, due to a similarly aligned local magnetic field. However, 
we find the apparent angular sizes on the average a factor of 2.5 less than those reported 
by Desai et al., indicating significantly less scattering than inferred earlier.
The average position angle of the scattered broadened images is also seen to deviate
significantly (by about 10 degrees) from that implied by the magnetic field in the Galactic plane. 
More intriguingly, for a few Zeeman pairs in our set, we find significant differences in the
 scatter broadened images for the two hands of polarization, even when apparent velocity 
separation is less than 0.1 km/s. Here we present the details of our observations and 
analysis, and discuss the interesting implications of our results for the intervening 
anisotropic magneto-ionic medium, as well as a comparison with the expectations 
based on earlier work.

\end{abstract}

\firstsection 
\section{Introduction}
W49N is a well-known and extensively-studied massive star-forming complex 
in the Milky Way. The OH, H$_2$O masers in this region represent some of
the most luminous of such sources in our Galaxy. 
In our efforts to study intrinsic short time-scale variability in W3OH (Ramachandran et al, 2006;
also Laskar et al, in this volume)
we were able to estimate and remove possible variability due to interstellar scintillation
using the fact that decorrelation bandwidth is larger than line-widths.
Such velocity-resolved
analysis of W3OH data has suggested intrinsic variability on 15-20 minute
time scale (Ramachandran et al., 2006), whereas W49N data show
variations on time-scales of 1 hour or longer (Goss et al., 2007, talk at IAUS 242, Alice Springs).

The W49N complex, in the 
Galactic direction  $l= 43^\circ.17$; $b=0^\circ.01$,
is located on the far-side of the Solar circle.
Its large distance ($\sim$11.4 kpc) and low Galactic latitude together
makes this object attractive for studying various propagation effects 
due to the intervening medium. The interstellar
scattering in this case is comparable to that in the Vela direction.
In an earlier study, Desai, Gwinn \& Diamond (1994; hereafter DGD94) have found significant
anisotropic scattering,
attributable to electron-density irregularities that are preferentially elongated 
in the Galactic plane.
In this paper, we describe our probe based on the observed manifestations of the 
interstellar scattering in the W49N direction, compare our results with those
from DGD94, and present intriguing cases of differential anisotropic scattering
apparent for the two hands of circular polarization.

\begin{figure}[b]
 \vspace*{-0.25 cm}
\begin{center}
 \includegraphics[width=2.9in,angle=-90]{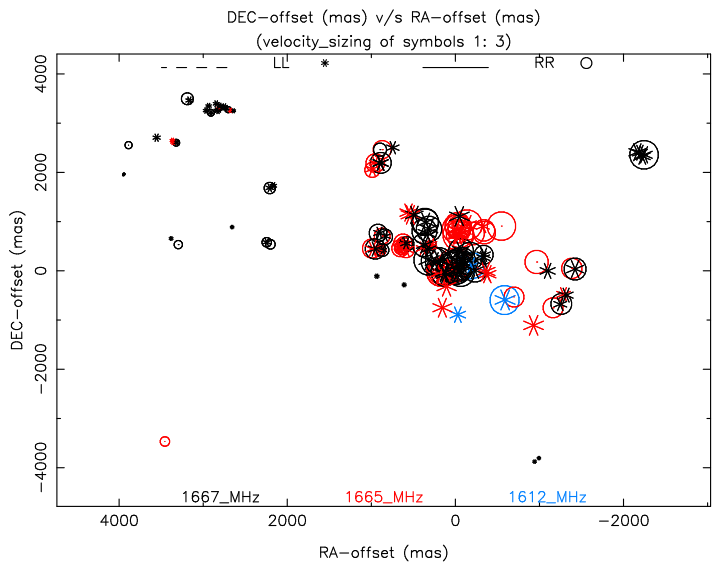} 
 \caption{The observed distribution of W49N OH maser sources in RA-Dec. The symbol sizes
are scaled proportional to the velocity (which are in the range +2 to +21 km/s). The velocity-position correlation, 
and the implied bipolar nature, is apparent from the majority of the sources in the sample.}
   \label{fig1}
\end{center}
\end{figure}

\section{Our data}
Our visibility data from 12+ hour synthesis observations with VLBA (BR107)
were calibrated \& processed in the standard way using AIPS.
The data were self-calibrated, and absolute coordinates are not available
due to an absence of phase reference source obervations.
The spectral-line image cubes, obtained separately at 1612, 1665 \& 1667 MHz,
provided high angular-resolution images of the set of sources for each
of the transitions, and for each of the circular polarizations. 
The imaged angular extent of about 8 or so arcseconds
corresponds to a transverse span of $\sim$~0.5 pc 
at the distance of the source, although a majority of
the sources are within central 2-arcsecond wide region. 
As a result of exclusion of the outer 2 antennas of VLBA 
from Self-Cal and this imaging, our synthesized beam size was compromised
to $\sim$20 mas and $\sim$15 mas in RA and Declination, respectively 
(corresponding spatial resolution being $\sim$200 AU and $\sim$150 AU).
Each data set consisting of cleaned and restored images across 
240 spectral channels spanning 22 km/s
velocity range, providing velocity resolution of $\sim$0.1 km/s, was
used to identify discrete maser sources (avoiding cases with
significant velocity gradients). For each of 
clearly identifiable discrete (or isolated) maser sources,
JMFIT-based estimates of 
the deconvolved source shape and size were obtained, 
in addition to the estimates
of the mean location and velocity, along with estimates of the 
respective uncertainties.
The resultant data on a total of 205 sources 
(most of these at 1665 \& 1667 MHz) were examined among
other things, for positional proximity of LCP and RCP source pairs ($\le$ 10 mas),
and a few dozen Zeeman pairs were thus identified.
Figure 1 shows the distribution of all of the discrete sources
in the RA-Dec plane, where the size of the symbol is proportional to
the mean velocity associated with the observed line.

\begin{figure}[b]
 \vspace*{-0.25 cm}
\begin{center}
 \includegraphics[width=3.8in]{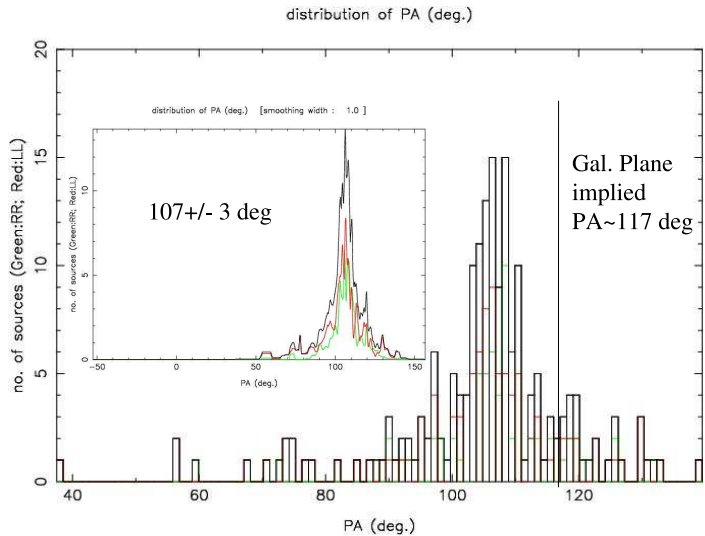} 
 \caption{Distribution of the position angles (PA) of the scatter-broadened images. The main panel
shows the histogram of the PA values for the set of OH maser sources in W49N. The vertical line
at 117 degrees, shown for reference, marks the mean PA expected from the anisotropic scattering 
due to electron density irregularities, if these were to be elongated parallel to the Galactic plane. 
The inset shows the distribution of PAs after accounting for uncertainties in individual measurements 
(green and red profile correspond to LCP and RCP, respectively). 
This distribution is used to
estimate the mean PA of the observed scatter-broadened images (107$\pm$3 degrees).}
   \label{fig2}
\end{center}
\end{figure}

\section{Anisotropic scattering: apparent source sizes and shapes}

Scatter-broadened shapes and orientations of the maser spots provide
valuable information on the nature of the scattering medium.
The elongations or the axial ratios (ratio of major to minor axis
of the best-fit ellipse) are found in the range 1.5 to 3, indicating
significant anisotropic scattering, consistent with the finding of
DGD94. However, we find the overall apparent
sizes of the sources to be significantly smaller (by a factor of $\ge$ 2)
than those reported by DGD94. Our size estimates are nonetheless
consistent with the suitably scaled values of the scatter-broadening 
of H$_2$O masers reported by Gwinn (1994).
Possible reasons for what appears to be an overestimation of the sizes in DGD94
are unclear, though remain intriguing. 
One of the most important aspects discussed by DGD94 (also see references therein)
was that 
the apparent image shapes resulting from  anisotropic diffraction
are expected to have elongation orthogonal to that of the 
scattering irregularities.  A presence of a magnetic field would
induce such anisotropy in the electron-density irregularities;
then the implied field direction would be perpendicular to
the resultant image position angle.
Based on their
limited sample, of the apparent image shapes for 27 spectral components from 6 sources,
DGD94 had suggested that the apparent 
anisotropy is induced by a magnetic field in and parallel to
the Galactic plane.

In Figure 2, we present a distribution of our estimates of 
the position angles for our significantly ($\sim$ 30 times) larger sample.
Some random spread in PA is apparent in both, our
and DGD94's, reported PA values, and such a spread is not unexpected.
However, we find that our PA distribution has a significant offset from
the PA of $\sim$117 degrees, the expected PA if the density
irregularities were to be elongated parallel to the Galactic plane.
Our sample gives a mean PA of 107$\pm$3 degrees, implying 
a significant mean deviation, of $\sim$10 degrees. The DGD94 sample
may have been too limited to make such a offset detectable.
What might be the reason for this systematic offset ?
In this context, a closer look at the magnetic field structure along
the W49N direction would be instructive. 
Interestingly, our sight-line to W49N is through 
the well-known North Polar Spur feature.
From the study, by Wolleben (2007) and others, of the North Polar Spur,
it is evident that
significantly different magnetic field structure as well as
enhanced scattering would be
expected for the medium within half a kpc of the Sun.
We note that the field orientation in this region, corresponding to the NPS,
might be at a large inclination, if not almost orthogonal,
to the Galactic plane. Also, given that the scattering medium closer to 
the observer is expected to
make relatively higher contribution to the angular broadening 
(e.g., Gwinn et al. 1993, Deshpande and Ramachandran 1998), 
the above mentioned PA deviation can be caused by the more local scattering
in the NPS region, with density anisotropy at a possibly large angle with respect 
to the Galactic plane. Careful modelling of the magneto-ionic medium would
reveal the relative strengths and sense of anisotropic scattering, implied
magnetic field strengths, for the different regions along the sight-line. 
Details and results of such modelling will be discussed elsewhere.

\section{Differential Anisotropic scattering: probe with Zeeman pairs}

\begin{figure}[b]
 \vspace*{-0.25 cm}
\begin{center}
 \includegraphics[width=2.1in,angle=-90]{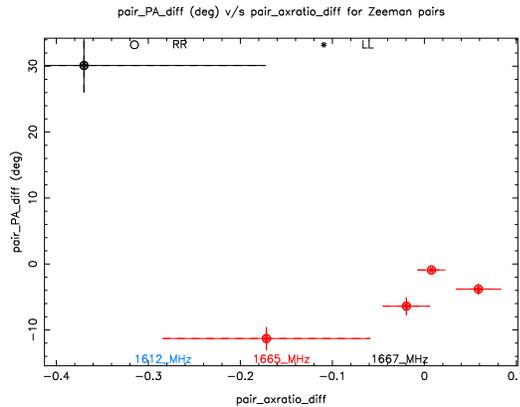} 
 \caption{The PA differences within Zeeman pairs versus the respective differences in axial ratios
for the cases where significant PA differences are apparent. The error-bars indicate
$\pm1-\sigma$ uncertainties.
All the axial ratios are all consistent with equal values for the two hands of circular polarization.  }
   \label{fig3}
\end{center}
\end{figure}

As mentioned earlier, a few dozen Zeeman pairs are identified from 
our large sample of sources, using positional-proximity criterion.
These Zeeman pairs allow us to probe yet unobserved aspect of scattering
caused by the magneto-ionic component of the interstellar medium.
A magneto-ionic medium would, in principle, render different refractive indices 
for the two hands of circular polarization.
Diffractive scintillation and scatter-image shapes should 
therefore differ for LHC, RHC due to line-of-sight component of the magnetic
field in the intervening medium.
Hence scattering-dominated images of even a randomly polarized source might 
show circular polarization in unmatched parts of the images, when the
Faraday rotation is significant.
Macquart \& Melrose (2000) indeed consider this possibility, 
but estimate the effect to be too small (circular fraction $\sim 10^{-8}$) to be observable.
Contrary to that expectation, the scatter-broadened images of some of the W49N 
OH maser sources seem to significantly differ in L\&R-hand circular polarizations,
i.e. within a given Zeeman pair !
For the rest of the Zeeman pairs, the PA differences are either small 
or within the respective uncertainties.
Figure 3 shows the observed differences in the image shape parameters for
a subset of our sample of Zeeman pairs. The subset includes only those cases for which
significant difference ($\ge$ 6$\sigma$) in image PA is observed, when compared 
to corresponding uncertainty $\sigma$. Although there are only a few such Zeeman pairs, 
the PA differences range between 6 to 30 degrees.
Difference in the line-velocities within most of these pairs, and hence in frequencies, 
is too small to account for the apparent differential scattering.
Position differences are also within a few mas, and are unlikely to 
contribute to the observed PA differences.

To explore the issue further, we have made  preliminary attempts
to simulate magneto-ionic medium with a mild anisotropy. 
A random column-density distribution of free electrons,
following a power-law spatial spectrum (with Kolmogorov index $-11/3$), is used
to simulate a 2-d scattering screen across a transverse extent of a few Fresnel scales.
For simplicity, a uniform magnetic field is assumed, so that
the phase screens for the two circular polarizations are merely scaled
versions of each other.
The net circular polarization, if
viewed with coarse resolution would be negligible, consistent with
the conclusion of Macquart \& Melrose (2000).  
However, when viewed with 10 or so milliarcsecond resolution, these preliminary
simulations do indeed reveal noticeable differential diffractive effects, and hence
the differing shapes and sizes of scatter-broadened images, 
for the two circular polarizations. 
Of course, more detailed simulations, with thick screens, will be needed
to assess these issues in detail.

\section{Summary}

The OH maser sources in W49N do show anisotropic scattering, 
but the apparent scatter broadening is much less than reported earlier.
The position angles of the source images deviate significantly from the value 
expected if scattering density irregularities were to be ``stretched" 
due to magnetic field strictly aligned parallel to the Galactic plane,
indicating significant scatter-broadening contribution from
differently aligned density irregularities, possibly
associated with the North Polar Spur. 
Some of our data also reveal differential scattering during propagation 
of the two circular polarizations, caused most likely by Faraday rotation.
Through various aspects discussed above, the attractive, but yet unexplored, 
potential of the high-resolution maser observations for 
probing the intervening magneto-ionic medium is certainly evident.

{\bf Acknowledgements:} It is a pleasure to acknowledge the contributions from R. Ramachandran
and Sarah Streb at different stages of the reported work. The National Radio Astronomy Observatory 
is a facility of the Natiional Science Foundation operated under a cooperative agreement 
by Associated Universities, Inc.

\end{document}